\documentstyle[useAMS,graphicx,longtable]{mn2e}

\newcommand{\hMpc}{{\ifmmode{h^{-1}{\rm Mpc}}\else{$h^{-1}$Mpc}\fi}}
\newcommand{\hkpc}{{\ifmmode{h^{-1}{\rm kpc}}\else{$h^{-1}$kpc}\fi}}

\def\approxlt{\mathrel{\spose{\lower 3pt\hbox{$\sim$}}
        \raise 2.0pt\hbox{$<$}}}
\def\approxgt{\mathrel{\spose{\lower 3pt\hbox{$\sim$}}
        \raise 2.0pt\hbox{$>$}}}
\def\approxpropto{\mathrel{\spose{\lower 3pt\hbox{$\sim$}}
        \raise 2.0pt\hbox{$\propto$}}}

\title{Automated classification of variable stars for ASAS~1-2 data}
\author[L. Eyer \& C. Blake]{L. Eyer$^{1, 2}$ \& C. Blake$^1$\\
   $^1$ Princeton University Observatory, Princeton, NJ 08544, USA\\
   $^2$ Observatoire de Gen\`eve, CH-1290 Sauverny, Switzerland}
\date{Accepted --.
      Received -- ;
      in original form --}

\pagerange{\pageref{firstpage}--\pageref{lastpage}}
\pubyear{2004}

\begin{document}

\maketitle

\label{firstpage}

\begin{abstract}
%===============

With the advent of surveys generating multi-epoch photometry and the
discovery of large numbers of variable stars, the classification of
these stars has to be automatic. We have developed such a
classification procedure for about 1700 stars from the variable star
catalogue of ASAS~1-2 (All Sky Automated Survey, Pojma\'nski 2000) by
selecting the periodic ones and by applying an unsupervised Bayesian
classifier using parameters obtained through a Fourier decomposition
of the light curve. For irregular light curves we used the period and
moments of the magnitude distribution for the classification.
In the case of ASAS~1-2, 83\% of variable objects are red giants. A
general relation between the period and amplitude is found for a large
fraction of those stars.
The selection led to 302 periodic and 1429 semi-periodic stars which
are classified in 6 major groups: eclipsing binaries, ``sinusoidal
curves'', Cepheids, small amplitude red variables, SR and Mira stars.
The type classification error level is estimated to be about 7\%.

\end{abstract}


\section{Introduction}
%=====================
\label{sec:int}

The knowledge of the bright sky variability is relatively poor. Bohdan
Paczy\'nski (2001) denounces this situation vigorously, ``I think this
ignorance is inexcusable and embarrassing to the astronomical
community''. Yet, at magnitude 12, it is estimated that $90\%$ of the
variables are unknown (Paczy\'nski 2000).

In recent years, only the {\sc hipparcos} satellite has done a
multi-epoch photometric all-sky survey with an associated analysis
aimed at systematically detecting variability.  This survey has a mean
of 110 measurements per star over 3.3 years.  It goes down to V
magnitude 7.3/9.0 depending on the colour of the star and its ecliptic
latitude $\beta$.  The result of this analysis for the main mission is
published in volumes 11 and 12 of ESA (1997) (see also the flags H6,
H49 to H53 of the main catalogue) and concerns 11562 variable stars
that comprise about 10\% of all catalogue entries. {\sc hipparcos}
enables the description of the behaviour of individual stars by giving
mean color and magnitude, parallax, period, amplitude and epoch of the
minimum or maximum light, but also for the stellar variability across
the HR-diagram (Eyer \& Grenon 1997) to be described globally.  It is
worth noting that the data classification in the different variable
types was ``manual'' and incomplete. The {\sc tycho} photometric data,
a deeper survey, caused more problems in its analysis and gave
somewhat disappointing results even after efforts to censor poor
quality data points (Piquard et al.  2001).

There are photometric optical surveys which have specific goals, like
OGLE, EROS, MACHO for detecting microlensing events, ROTSE and LOTIS
for detecting gamma ray bursts. We don't discuss those here and refer
to Paczy\'nski (2000). They were exploited for other purposes and
global analyses have been done (Belokurov et al. 2003) or are underway
(Wo\'zniak P. et al. 2001).

There are many ways to conduct a survey since several competing
parameters cannot all be maximized. Four primary questions are
in competition: how deep, how frequently sampled, how wide, and how
precise is a survey?  Consequently, different astronomical
subjects/objects will be explored/discovered depending on these
choices.

% and we address the question of classification of the stars of that
% selection.

It is important to consider this work in view of the general research
trend. With surveys from the ground and from space going wider and
deeper, the number of objects is increasing dramatically and the
handling of data is becoming more difficult.  With exponential data
growth, there is a clear need for automated algorithms.

We are analysing here the ASAS survey data obtained during its test
implementation (phase 1 and 2).  ASAS is described in the next
section. An extraction of the variable objects was done by Pojma\'nski
(2000) and we address here the question of classification of the stars
of this selection. The number of objects that are considered is about
3900 variable objects, among them about 400 periodic variable objects
(Pojma\'nski 2000).
% 3884 objects

Pojma\'nski (2002, 2003, 2004) has continued to develop ASAS, and
presented an analysis of the third phase of ASAS. About 1.3/3.2
million stars brighter than $V=15$ were measured, 3126/10453 variable
stars were extracted and classified as eclipsing (1046/1718), regular
pulsating (778/731), Mira (132/849) and ``other'' (1170/7155), mostly
SR, IR and LPV stars (the numbers refers to the year up to 2003/2004).
The regular pulsating stars have been separated into $\delta$ Scuti
stars, RRab and RRc stars, $\delta$ Cep stars (fundamental and first
overtone pulsators). Pojma\'nski's classification methods differ from
the method presented here. In his article of 2003, he used carefully
selected two-dimensional projections of the Fourier coefficient space
where the separation of variable types are pre-analysed and well
distinguishable.  This method was used by Ruci\'nski (1993) to
distinguish between contact binaries and detached ones. In
Pojma\'nski's study of 2004, he added 2MASS photometry, using two
additional parametric planes ($H-K$,$\log P$) and ($H-K$,$J-H$).

%Given the data size of ASAS, the analysis is still manageable in a
%non-automated way. This permits a thorough verification of the
%procedures and allows the correct and incorrect classification rates to
%be checked. However, this procedure could be implemented in any large
%surveys with random manageable subsamples.

The methodology presented here is divided in three main objectives: 1) to
search for periodicity, 2) to model the light curve and characterize
the data by a set of parameters and to remove dubious cases, 3) to
determine the variability types.

%The goal is to implement such a method in different surveys and to
%test them against other method (). This is therefore the first attempt
%to have a global approach to a database.

%The precision of ASAS renders the task easier with respect to other
%surveys like OGLE or astrometric missions

%The test sample of ASAS seems ideal... size... complexity
%precision...
%To put in context.
%The methodology which was advocated by Paczy\'nski, applied in OGLE and
%followed in ASAS are to start by small size projects  and then to
%increase gradually complexity. This is also the path followed here.
%This work aims at being able to handle a large amount of data, to
%process it and classify it. The ASAS test-data gives a first step in
%that direction.

%The future.
%extremely accurate which will permit to explore space parameters.

%This work is also permitting us to gain expertise in the field of
%classification. The forthcoming

\section{Data Description}
%=========================
\label{sec:datadesc}

ASAS (All Sky Automated Survey) is a photometric survey. Its goal is
to regularly monitor the sky so as to detect any variable phenomenon.
In its testbed, during the years 1997-2000, ASAS~1-2 repeatedly
measured 50 fields (2x3 deg$^2$ each) and obtained photometry for
about 140'000 stars in I-band with a 135 mm photolens and a 768x512
pixel CCD camera.  Among the 140'000 observed stars, a set of 3890
variable objects was extracted by Pojma\'nski (2000).
%The size of the psf is , the size per pixel is.

The limiting magnitude is about $I=13$ and saturation occurs for stars
brighter than $I=7$.  The precision of the measurements is 0.01
at 8 mag and degrades to 0.07 at 12. Those estimates were derived from
the difference of successive measurements (the underlying assumption
being that the variability is negligible for short time scales, see
Eyer \& Genton 1999). The quoted individual photometric errors are not
always fully consistent with our estimates.

\begin{figure*}
\includegraphics[width=120mm,angle=-90]{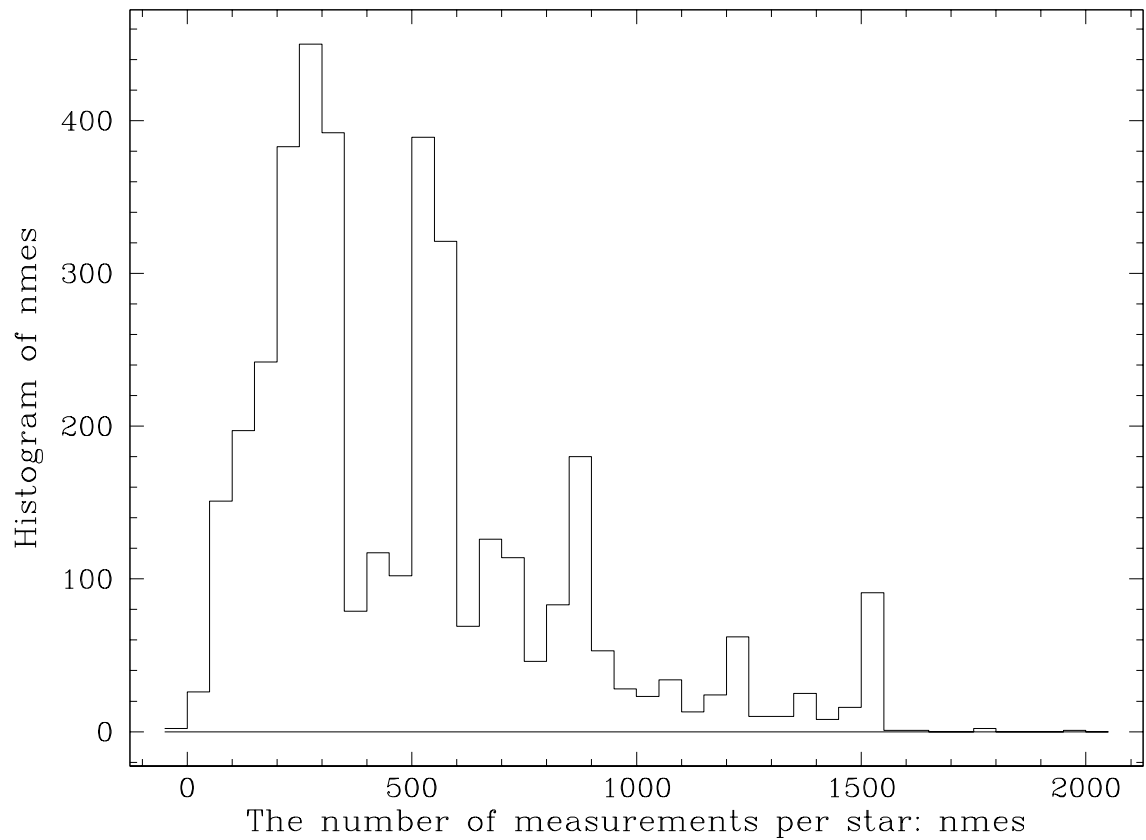}
\caption{\label{fig:histonmes}
 The histogram of the number of measurements per star.
}
\end{figure*}

\begin{figure*}
\includegraphics[width=120mm,angle=-90]{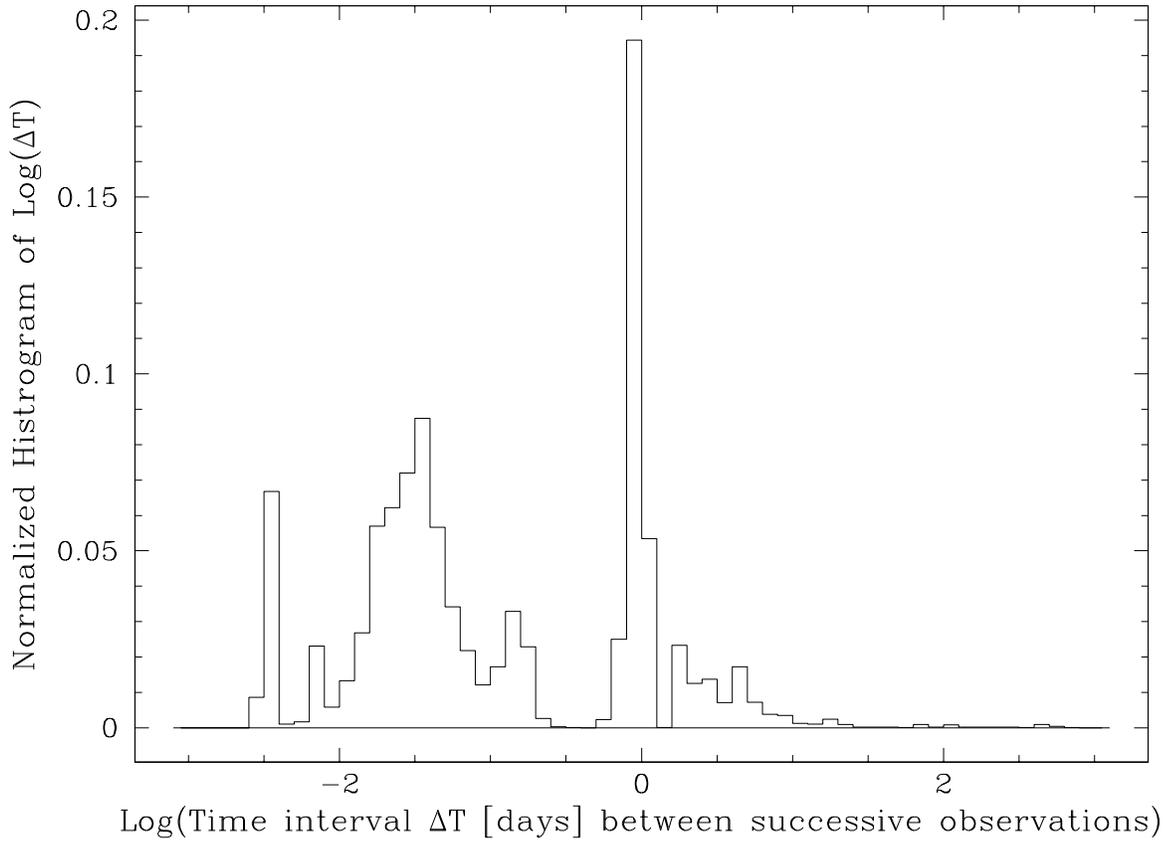}
\caption{\label{fig:norhis}
The normalized histogram of time intervals between successive
measurements.  We can see the regular nightly observing
pattern. However, there are other characteristic time intervals for instance
$\sim$5 minutes or $\sim$45 minutes.
}
\end{figure*}

The dates are given in truncated Heliocentric Julian dates
HJD-2450000.  The time sampling and number of measurements are quite
diverse. There is a large gap between 585 HJD-2450000 and 1020
HJD-2450000 which was caused by a flood in the shelter containing the
telescope and the control apparatus. The histogram of the number of
measurements per star is given in Fig.~\ref{fig:histonmes} and the
histogram of time differences between successive observations in
Fig.~\ref{fig:norhis}.

\begin{figure*}
\includegraphics[width=140mm,angle=0]{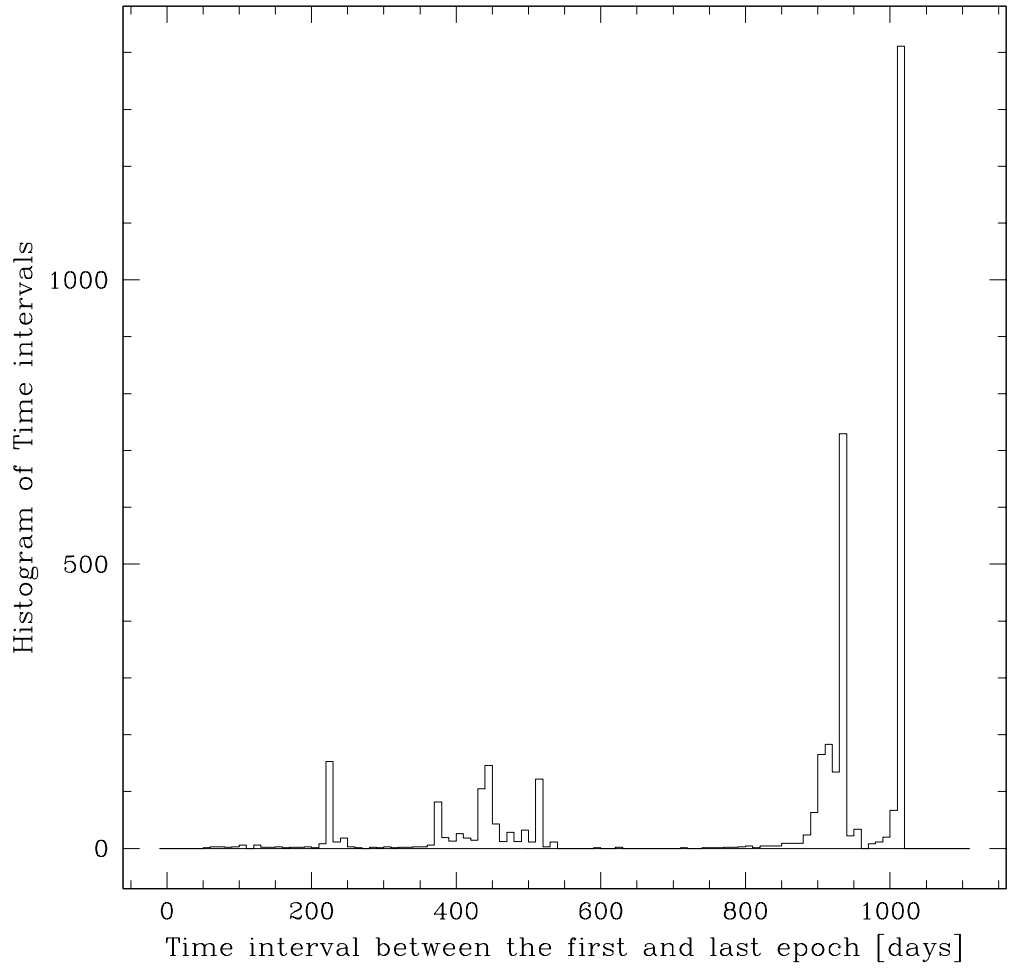}
\caption{\label{fig:deltaT}
Histogram of the difference between the last and first epoch. The star
sample is somewhat heterogeneous, but with many stars being
covered over a period of nearly 3 years. However, those stars may have
large intervals without observations.
%In fact 2.8 years for the longest peak
}
\end{figure*}

For our analysis, extreme values, which were more than 4 times the
dispersion distant from the mean, were removed.  Different fields had
different time coverage and time span.  This data heterogeneity
complicates the analysis. Fig.~\ref{fig:deltaT} shows the time
difference between the last and first epoch for each star. Since large
gaps might cause problems in the analysis, only observations after
1020 HJD-2450000 were considered.  However, the data before 585
HJD-2450000 were used to verify shorter periods.

% The total number of observations removed amounts 23%, note
% the different number given by Cullen in his email.

\section{Period search}
%======================
\label{sec:persea}

% Programme classif1.f
The data were processed through the Lomb (1986) period search
algorithm, which is equivalent to adjusting the parameters of a
sinusoidal curve.
We used the algorithm given by Press et al. (1992), in its fast
version called ``fasper''. The algorithm has two parameters which fix
the resolution in frequency and the highest frequency searched.  The
first, $\mbox{ofac}$, was set to 6 while the second,
$\mbox{hifac}$, was changed according to the properties of the light
curve and of the sampling. In fasper, the highest frequency searched
is computed given the time intervals, the number of data points and
the ad hoc factor multiple ($\mbox{hifac}$) of Nyquist frequency as
defined by Press et al. (1992). This factor is needed since short
periods can be detected even if the data has large time gaps between
measurements (see Eyer \& Bartholdi 1999).  We did not use a constant
hifac parameter, since in certain cases (large amplitude variables) it
is unnecessary to search for high frequencies. On the other hand, for
short time variability very high frequencies can be found.

We computed $\sigma_N$, the dispersion of the difference of
measurements separated by a small time interval (less than 1.5 days),
and divided it by the dispersion of the signal. If this ratio is small
($< 0.55$), we used hifac such that the highest frequency is equal to
0.5 1/day, if it is between 0.55 and 0.66 we used $hifac=9$, and if
above 0.66 we fixed hifac so that the highest frequency is equal to 17
1/day. This was found to be a better compromise than using a single
value for hifac. Long time scale and large amplitude variable stars
can show a variety of irregular behaviours and trends, and therefore
are prone to aliasing.  The determination of the highest frequency to
be searched was limited to 17 1/day. This limit was determined for the
shortest pulsators which have high amplitudes ($> 0.05$ ) using the
$\delta$ Scuti star catalogue of Rodriguez et al.  (2000). The choice
of $\delta$ Scuti stars is motivated by the fact that they have the
shortest periods among the high amplitude periodic variable stars.
Furthermore, there is a large number of known candidates allowing
statistical conclusions to be drawn.  Only a small fraction of those
stars would be overlooked if we take the limit at 17 1/day.
% set to 10% (or 5% same result) see the /u/leyer/SMONGO/delscuti_pa

The frequency of the highest peak  in the Lomb periodogram is selected
as being the   main frequency of the  signal.  The amplitude $A$ of  a
sinusoidal   signal  with   a  standard  deviation  of   $\sigma$  and
$\mbox{nmes}$ measurements is related to the power $P$ by the relation
$A=\sqrt{4 \, \sigma \, P/\mbox{nmes}}$.

%%\section{Selection of subsample}
%%%==============================
%%\label{sec:sel}

%%We apply several criteria to remove dubious periods in order
%%to have a subsample of sufficient quality for classification.

%%\begin{itemize}
%% \item The stars with less than 40 measurements are disregarded.
%%%       apparently 6 stars were rejected
%%  \item A parabolic curve is fitted to the data, then subtracted
%%        to it and the Lomb algorithm is recomputed on the resulting
%%        data under the same conditions than the initial trial described
%%        above. We present the diagram of
%%        the initial period and the period found after having subtracted
%%        that parabolic fit in Fig.~\ref{fig:perperd}. We remark that
%%        one-day and half-day spurious periods are present.
%%        We also remark that there are periods which are not stable.
%%	All those cases are rejected.
%% %  \item For $\sigma_N > 0.55$, we compute the Lomb periodogram for the
%% %        data before $HJD-2450000 =900$ if there are more than 40 data points,
%% %        [NOT extremely useful\ldots].
%% %  \item variograms
%%   \item A linearisation of the Fourier series is done, to study the
%%         convergence. When the convergence is not attained within a reasonable
%%         frequency interval, the star is rejected from the sample.
%% \end{itemize}
% Eliminer les lorsque les periodes sont  plus grande que l'intervalle
% de temps des mesures pour l'etoile donnee. Ici nous allons faire par
% groupe de points. Autreme

\section{Fourier series}
%=======================
\label{sec:four}

% Programme  classif2.f
A Fourier series with $n (\geq 2)$ harmonics is fitted to the data,
$$
S(t)= \sum_{i=1}^{n} A_i \sin(2 \pi \nu t) + B_i \cos(2 \pi \nu
t),$$
$\nu$ being the frequency, $t$ the time. The parameters
$$R_{21}=\sqrt{\frac{A_2^2+B_2^2}{A_1^2+B_1^2}}$$ and
$$\phi_{21}=\arctan{(-A_2/B_2)} - 2 \arctan{(-A_1/B_1)}$$
are computed. We linearize the equations with respect to the
frequency, and search for the least square fit iteratively. The
initial frequency is the one obtained with the Lomb algorithm.

The determination of the number of harmonics was done
iteratively. Initially, the number of harmonics is fixed at 2, so that
$R_{21}$ and $\phi_{21}$ are always defined.  The procedure is then to
loop and stop at a maximum of 6 harmonics. We determine a first
solution for a certain number of harmonics, then we increase by one
the number of harmonics and recompute a second solution. We perform a
Fisher test comparing the two models (Lupton 1993). If there is a
significant reduction of the $\chi^2$ by adding a harmonic we repeat
the procedure by adding one more and if not we keep the first
model. The majority of stars (58\%) have a solution with two
harmonics, then 18, 11, 7, 6 \% have solutions with 3, 4, 5, 6
harmonics respectively.

\subsection{Estimation of the Error on the Period}
%-------------------------------------------------
\label{sec:errper}

The general trend is that the error on the period is function of the
square of the period.  However, peculiarities in the light curve such
as sharp features, like rising branches in Cepheids, may fix the period
more precisely.

The method used for estimating the error on periods is the same as the
one used for the {\sc hipparcos} catalogue (ESA 1997) in the case of the
Geneva solutions (Eyer 1998). We use the estimation of the error on
the frequency given by the least square fit of a linearized Fourier
series.
Schwarzenberg-Czerny (1991) proposed an estimate for the error on the
period by taking into account the correlation of the residuals. Such
correction is not implemented here.

%\subsection{One day alias}
%-------------------------
%To remove with variogramme (Eyer \& Genton 1999).

\section{Cleaning the sample}
%----------------------------
\label{sec:cleansamp}

\begin{figure*}
\includegraphics[width=140mm,angle=0]{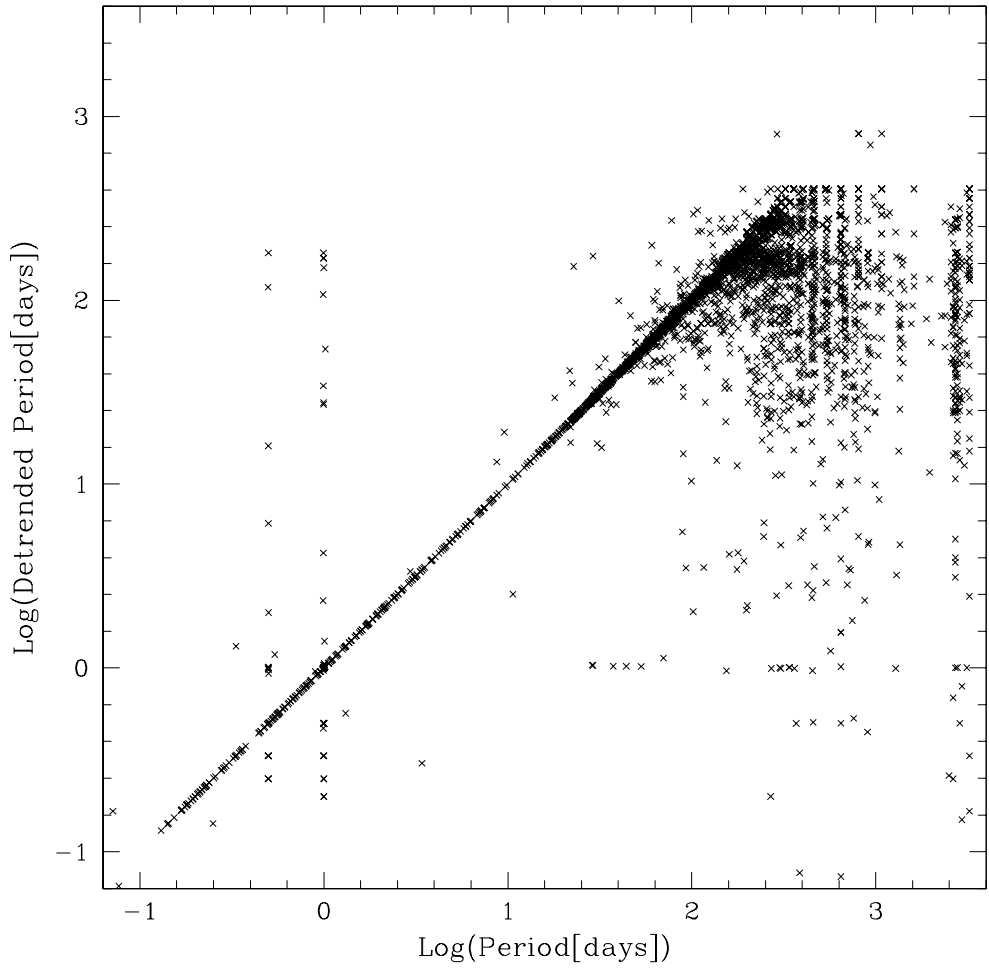}
\caption{\label{fig:perperd}
Comparison of the periods, obtained by the Lomb algorithm on the data
and when a parabolic fit is subtracted from the data, allows
some aliases as well as unstable solutions to be identified.
}
\end{figure*}

To select a ``well behaved'' data set, we used the following criteria:

\begin{itemize}
 \item The stars with less than 40 measurements are discarded.
%  apparently 6 stars were rejected
 \item We defined a reduced time span: The total observational time
   span where the three largest gaps are subtracted. We retained
   periods which are smaller than the reduced time span divided by
   1.2.  It was found empirically that many short frequencies are
   spurious.  Such a criterion may exclude true Long Period Variable
   stars with poorly covered light curve.
%   per < delT - (tij1+tij2+tij3 )
 \item We rejected objects with a skewness on the I~magnitude larger
   than $-1$ (rejecting time series containing bright outliers like
   flares, cosmic rays).
 \item We selected objects which have a mean I~magnitude smaller than
   $12.65$.
 \item A parabolic curve was fitted to the data, then subtracted from
   it and the Lomb algorithm was recomputed on the resulting data
   under the same conditions as the initial trial described in
   section~\ref{sec:persea}.  We present the diagram of the initial
   period and the period found after having subtracted that parabolic
   fit in Fig.~\ref{fig:perperd}. We notice that one-day and half-day
   spurious periods are present.  We also notice that there are
   periods which are not stable.  We selected the objects which have a
   difference of the two determined frequencies that just allows a
   confusion between neighbouring peaks in the Fourier spectrum.
%   We define abs((nu-nu_d)/(\delat f)) and select this quantity to be
%   smaller than 15.
 \item Periods near the aliases of one day and half a day were
   removed.
\end{itemize}

These criteria were established mostly empirically after the samples
rejected were studied in a detailed manner to avoid rejecting
valuable objects.

\section{Autoclass}
%==================
\label{sec:aut}

Humans by nature have trouble visualizing multidimensional data sets
especially those with more than 3 dimensions.  More than 3 attributes
are needed to classify the light curves by the proposed Fourier
decomposition.  Furthermore, the variable star population is diverse.
Some groups have very well defined characteristics, others have
overlapping properties.  Some classes are divided almost arbitrarily
into different groups although they represent a continuum. For
example, as defined in the GCVS (Kholopov et al. 1985), Mira stars
have a peak to peak amplitude in V larger than 2.5 mag.  An
unsupervised program might give some indications for better divisions.
Another advantage of such an unsupervised algorithm is that it can
point out new classes of objects.

Autoclass (Cheeseman, Stutz 1996) is a Bayesian classifier.  The
algorithm looks for the number of classes and the classification which
is most probable, given the observed data.  The method was successfully
applied to several astronomical sets: {\sc iras} sources (Goebel et al.
1989), asteroids (Ivezic et al. 2001), and {\sc hipparcos} data (Eyer
unpublished).

The method is not fully automated since there is an interactive part
(the class models have to be specified), but the method takes the data
in its totality and proposes a broad classification.  As pointed out
by the authors of Autoclass, this interactivity is necessary.

Another useful aspect of the Bayesian classifier is that it computes a
class membership probability. Therefore, this probability can be used
to sort according to reliability levels within the classification.

The attributes (i.e. the parameters) chosen for the classification are
the Period, Amplitude, Phase Difference $\phi_{21}$ and Amplitude
Ratio $R_{21}$.  With only these four parameters we show that we reach
a rather reliable classification for the clear variables. For the
irregular variables we used the period, second, third, and fourth
moments of the light curves as the classification parameters.

As period and amplitude are positive quantities, we choose the
logarithm of those values for the classifier.

The phase difference $\phi_{21}$ is defined modulo $2\pi$. Circular or
angular real valued attributes are not yet available in Autoclass.  The
eclipsing binaries which constitute a major part of the sample we
classify have a period which is generally wrong by a factor of 2
(typical for Fourier type methods).  This means $\phi_{21}$ is often
around zero.  Therefore, the eclipsing binaries will be split in two
groups (those with a $\phi_{21}$ above $0$ and those below $2\pi$),
giving rise to many more classes.  For this reason, we redefined the
phase difference $\phi_{21}$ as being between $\frac{3\pi}{4}$ and
$\frac{3\pi}{4}+2\pi$. Few were found to have a $\phi_{21}$ around
$\frac{3\pi}{4}$.

There is a relation found between period and amplitude (see next
section).  The stars forming this relation have irregular light curves
and so the parameters have a large enough dispersion to weaken the
abilities of Autoclass to perform the classification.  We previously
(Eyer \& Blake 2002) selected a subsample of those objects manually,
to retain only fairly well behaved objects.  Here, we want a more
automated process. Instead of finding a method for selecting objects
with high residuals, we just divided the sample in two with the
relation $\mbox{Amplitude}=10^{-3.2} \mbox{period}^{1.6}$.  We then
apply Autoclass to these two samples with different attributes.

In our experience, adding parameters often does not improve the
classification. So the general method is to start with very few
parameters, apply Autoclass and analyse the result of the
classification (on a sub-sample for example).  If well known classes
are not separated we can add a parameter and iterate the process.

% Question: 
% --------
% the posterior probability is maximised when
% the number of classes equals the number of individual (cases).
% Is it in the prior of the model that such cases are penalised ?

\section{Red giants and period amplitude relation}
%=================================================
\label{sec:res}

\begin{figure*}
\includegraphics[width=120mm,angle=-90]{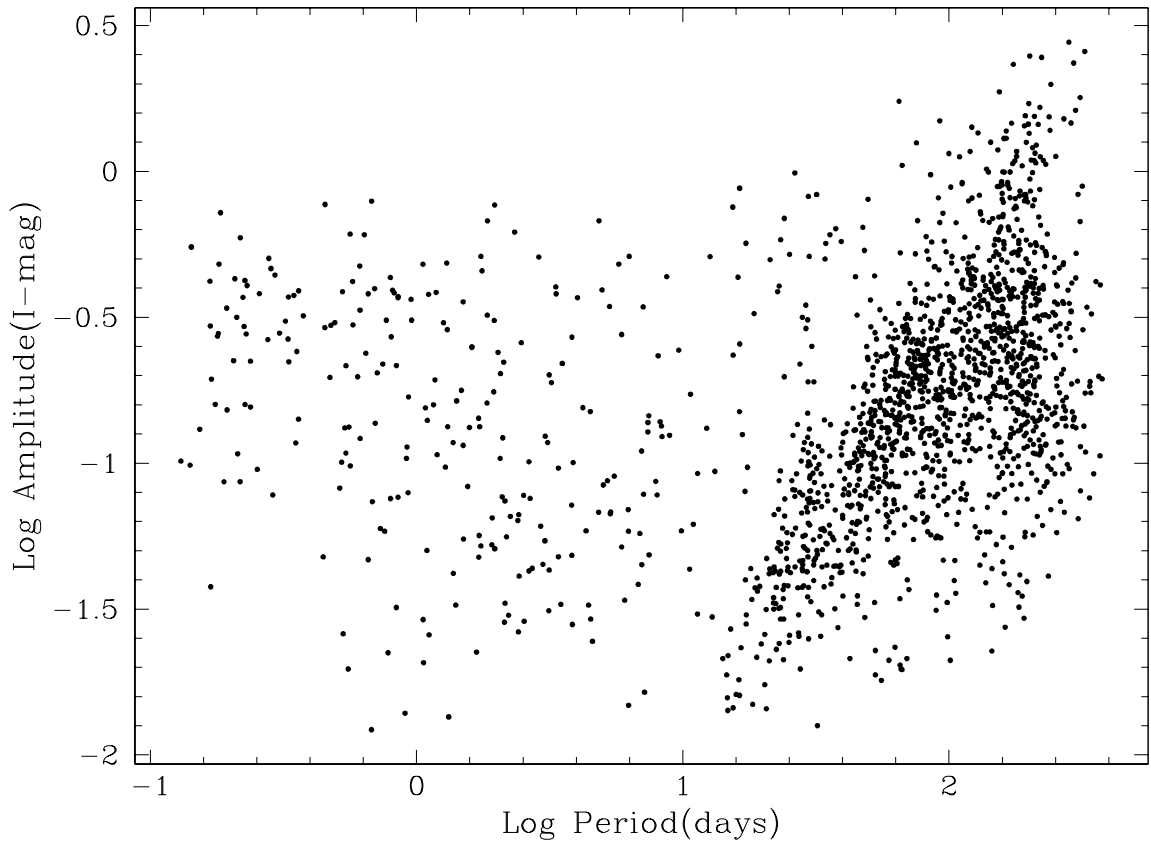}
\caption{\label{fig:lperlampli}
The raw diagram  $\log(\mbox{period})$,
 $\log(\mbox{amplitude})$. The diagram is dominated by red stars which
seem to fall on a relation (or several neighbouring sequences).
}
\end{figure*}

Fig.~\ref{fig:lperlampli} shows that there is a period
amplitude relation for a very large fraction of stars. We find that
about 83\% of the variables fall in this broad region (not
including Cepheids). If we compare the population of stars observed by
{\sc hipparcos} in that same region of the period amplitude diagram
(cf. ESA 1997 and Koen \& Eyer 2002), we find that most stars have
spectral types from K giants for the lower left part of the relation
to M giants for higher right. At the small amplitude and short
period side of this relation, we find the small amplitude red giants
studied for example by Percy et al. (2001).  At the large amplitude
and long period side, we find the well known Mira stars.
With the continuation of ASAS, more stars forming this relation will
be found, more data will be available per star, and the morphology of
this relation will be described with better precision. For the moment
we notice that this relation could also be formed by several adjacent
relations.

This period-amplitude relation is also observed in K-band infrared
photometry cf. van Loon (2002). Substructures and parallel relations
had also been observed by Minniti et al. (1998) and Wray, Eyer \&
Paczy\'nski (2004).

%This relation of period amplitude can be expected. If a group of star
%is pulsating in one identical mode, then at the boarder of the
%instability strip, the mechanism which excites the mode won't be as
%efficient as in the center of the strip and the amplitude will be
%smaller.

%We can also expect that the higher mass stars will have longer periods
%and also larger amplitudes as it is observed for the Cepheids.  So
%there is a mixture of these two effects.

%Gautschy: citer Gautschy ?

\section{Results of the Classification}
%---------------------------------------

%max_n_tries = 1000
%
%0           real        location       "per"                 error 0.009
%1           real        location       "ampl"                error 0.002
%2           real        location       "r21"                 error 0.002
%3           real        location       "phi21"               error 0.002
%4           real        location       "s3"                  error 0.005
%
%ignore 4
%# single_normal_cm 1  
%single_normal_cn 0 1 2 3

A prior work included 458 stars (Eyer \& Blake 2002), and now the
sample is extended to 1731 stars divided into two groups of 302 stars
and of 1429 stars. Thus 45\% of the stars in the sample have a
sufficiently regular behaviour to have been selected by our criteria.

\begin{figure*}
\includegraphics[width=140mm,angle=0]{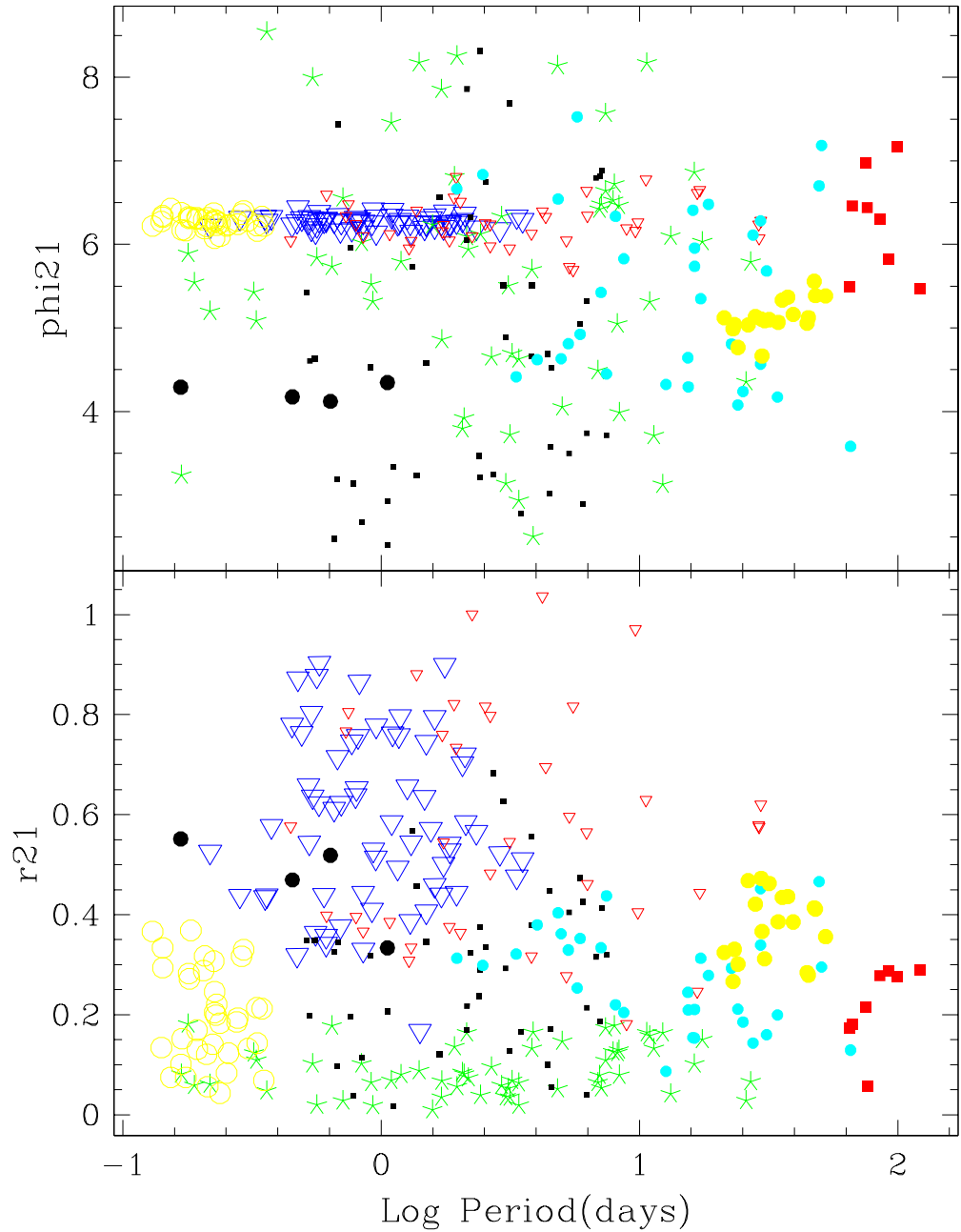}
\caption{\label{fig:lperphi21r21}
 The classification obtained in the diagram $\log(\mbox{period})$,
 $\phi_{21}$ and  $\log(\mbox{period})$, $R_{21}$.
 Symbols as in Fig.~\ref{fig:classpa}.
}
\end{figure*}

For the subsample of 302 stars, the stars are classified in 9 groups.
Certain groups appear to be very clean and others seem to contain more
difficult cases.  See Fig.~\ref{fig:classpa} and
Fig.~\ref{fig:lperphi21r21} for the result of the classification in a
$\log(\mbox{Period})-\log(\mbox{Amplitude})$ diagram and
$\log(\mbox{Period})-\phi_{21}$ or $R_{21}$ diagram.
We have:
\begin{itemize}
\item Eclipsing binaries ($\sim$192): One group with (63 stars)
  eclipsing binaries of EA and EB type. This group has no ambiguity of
  classification.  Another group (36 stars) of EW type eclipsing
  binaries, includes very few potential pulsating stars (like
  $\delta$ Scuti stars) or Ap stars with very sinusoidal curves. The
  third group (38 stars) contains more difficult cases, a large majority
  are eclipsing binaries, some seem marginal, and others have clearly
  a wrong period.
  %($\sim$144)
\item Cepheids ($\sim$19+13=32): We can find this type of variables in
  two different classes. One is very well defined (19 stars). Only one
  star seems to be peculiar in changing its amplitude. There are about
  13 other cases which could be Cepheids, some undoubtedly
  recognizable to human eyes, while others are difficult to recognize.
  %($\sim$48)
\item RR Lyrae stars ($\sim$ 4): probably one delta Scuti and 3 RR
  Lyrae of ab type, RR Lyrae of c types will be mixed with Eclipsing
  binaries of EW type.
\item LPVs (Long Period Variable stars): 8 stars are classified in a
  group with poorly defined light curves, and with periods above 60
  days. The time sampling is often sparse and does not cover many
  cycles of the light curve. The phase coverage often presents gaps.
\item Small amplitude variables ($\sim$ 44): many light curves seem to be
  marginal cases. Very few unambiguous cases could be identified.
  However, it makes sense to have such a group. There are some $\alpha$CVn
  which could be present in this group. The mean amplitude of this group
  is of 0.04 I-mag.
  %($\sim$100)
\item The last group ($\sim$ 58) is also composed of difficult cases   
  but of larger amplitude (mean amplitude is 0.12 I-mag) than the
  previous one. Here the variability is strongly detected.  It is worth
  noting that the formation of this group is a remarkable aspect of the
  classifier. Instead of spreading those objects among well defined
  classes, it is putting them in a separate group.
\end{itemize}

In total from this classification there are 5 groups out of 9 which
contain clear cases with an error of classification below 7\%, 2
groups, with some mixed classification, one group of small amplitude
variables which can be caused by many different effects, and one group
with very difficult cases.

The RR Lyrae stars, because of the limiting magnitude of the ASAS
survey, are too faint to be numerously detected.  Indeed the SDSS
survey (Ivezic et al. 2000) shows that the halo RR Lyrae stars are
very rare below I-mag 13.  Therefore, only 3-4 RR Lyrae stars are
found in the sample. The classification algorithm sometimes identifies
the RRab type, depending on the precision that we estimated for the
period, but these stars are not forming a stable class.  However, it
is remarkable that the program can form a new group with such a small
number of stars (about 1\% of the sample).  The classification was
found to be sensitive to error parameters and sometimes group RR Lyrae
are lost.

Beltrame and Poretti (2002) found that ASAS star 112843-5925.7
(HD304373) is a double mode Cepheid, the second one detected in our
Galaxy, pulsating in the first and second overtones or radial mode. In
our study, unfortunately the period search was limited for that star
to a frequency interval up to 0.5, missing the main peak of this short
period pulsator (Period $=$ 1.08~[1/day]). It is classified in the
group of stars which is a mix of pulsating and eclipsing.

The second group of 1429 stars, where we use the moments of the
distribution, is divided by Autoclass in 5 groups.  The classification
divides those stars into:
   SARV ($\sim$230),
   SR ($\sim 1158=102+484+572$),
   Mira ($\sim$41).

With these groups it is difficult to determine error levels since the
classification is extremely difficult to establish.  There are some
amusing case, the star ASAS180057-2333.8 is in the group of Mira
stars, but is clearly a long period eclipsing binary.

The catalogue, the light curves, and folded curves on individual basis and
on class basis can be seen at the first author's website.

%For an absolute magnitude of about 0.7 this correspond to a sphere
%of less than 100 pc.

%The level of false detection is established by inspection of
%subsamples, and is very variable but can be well below 7\% (TBC)
%for certain classes.

%The eclipsing binaries would be better classified using a principal
%component analysis, since the Fourier series is not optimal for
%describing the behaviour of those stars. It should be noted however
%that the subdivisions of the main classes has a tendency to separate
%the eclipsing binaries into EW, EB, EA types.

The eclipsing binaries have recomputed solutions where the initial
period is doubled since the Lomb-Scargle algorithm usually gives half
of the true period. The star 144245-0039.9 even has a factor 4 between
the true period and the period found by the Lomb algorithm.

\section{Conclusion}
%===================

We have developed a scheme for general and automated classification
for the periodic variable stars of ASAS~1-2 data set. Of course every
survey has its own properties, so even if the approach is transferable
in its principles, it probably requires modifications for every data
set. The general method, however, can easily be applied to larger
databases.

% It is
% manageable to study thoroughly several thousands stars, when the
% number of objects is getting higher it is possible to work with
% random subsamples to asses the quality of the analysis.

The work was broken into three parts: a) Selection of periodic
objects, b) modeling that subsample with Fourier series or with
simpler parameters like moments of the distribution and c) application
of the "Autoclass" Bayesian classifier.  At every step it is critical
to check the quality of the analysis, to interact with the data and to
visualize data or the defined parameters.

 Other studies are awaited (for example Wo\'zniak et al. 2001), using
different methods.  It is important to compare the efficiency of
the methods, their capability to detect new classes, their error
levels, and their CPU consumption. It might be interesting to apply
such a Bayesian classifier to the data of ASAS~3 and compare it with
the results of Pojma\'nski (2002, 2004).

%We would be very interested in studies where other algorithms are used
%as for example the neural networks and to compare such methods with the
%proposed Bayesian one.

At present, and in the very near future, there are several good
opportunities to apply such a classification method to other data
sets:
\begin{itemize}
 \item ASAS continues to survey the sky.
 \item The HAT project (Bakos et al. 2002) has released data
       (Hartman et al. 2004)
 \item The Magellanic Clouds data of OGLE-II (\.Zebru\'n et al. 2002)
  is available (58'000 variable stars), as well the 49 bulge fields
  from OGLE-II (Wo\'zniak et al. 2002). 
  The classification of bulge
  field 1 has been done by Mizerski and Bejger (2002), specific
  extractions of eclipsing binaries, RR Lyrae, Cepheids stars have
  been accomplished for the Magellanic clouds.
 \item The third phase of OGLE, OGLE-III, is functional and is taking
  data. The data rate is multiplied by a factor of 10 with respect to
  OGLE-II.
\end{itemize}

Real time detection and classification of phenomena such as supernovae
will require additional software development to be scientifically
valuable. For example, OGLE has possessed an Early Warning System
(EWS) since 1994 and received further development in 2003 to include
the detection of the effect of planets in a microlensing event.  It is
envisioned that the OGLE data will be put into the public domain
within 24 hours.  If so, OGLE-III opens the possibility for anyone to
make quasi-real time detection of variable phenomenon.

ASAS also has an Alert service. The photometric reduction pipeline is
available in real time within 5 minutes. The current service is
focused on the monitoring of cataclysmic variable stars.

Similar software has to be developed for large surveys from the ground
and from space like the GAIA mission (ESA 2000), so it is important to
gain knowledge of different classification methods.

%The detection and classification in real time is an other issue which
%is extremely important but not studied here. We think that methods
%should also be developed in that direction.

\section{Acknowledgements}
We are thankful for the help, comments and encouragements of
B.~Paczy\'nski, C.~Alard, A.~Gautschy, M.~Grenon, Z.~Ivezic and
G.~Pojma\'nski. Our thanks go to M.C.S.~Peterson for English corrections,
W.~O'Mullan and A.~Dartois for their help in Java programming.
The software SM (Lupton and Monger, 1997) was widely
used for this research.  Partial support for this project were
provided by the Swiss National Science Foundation, the NSF grant
AST-XXXXXX and the Carnegie Institution of Washington, Dept. of
Terrestrial Magnetism.

%References
%==========

%------------------------------TABLE 1-------------------------------------
\begin{table*}
 \caption{\label{tab1}
    Results of the classification (exctract, the full catalogue is available
    in the online version of the article). The columns are the ASAS ID (equatorial
    coordinates in equinox 2000), mean $I$ magnitude $\bar{I}$, standard error $\sigma_I$,
    the number of measurements $N$, the period, the amplitude, the amplitude
    ratio R21, the phase difference $\phi_{21}$, the number of harmonics $nh$,
    the classe $cl$, the probability of membership Prob.}
  \begin{center}
  \begin{tabular}{rrrrrrrrrrr} \hline
ASAS ID     &$\bar{I}$&$\sigma$& $N$ &    Period&   Ampl & $R21$ &$\phi_{21}$&nh&cl&Prob\\ \hline
$005759+0034.7$ & 10.444 & 0.028 & 1204 &   0.7980 &  0.076 &  0.396 &  6.242&  6 &3& 0.42\\
$015647-0021.2$ & 11.041 & 0.047 &  417 &   0.5427 &  0.108 &  0.102 &  7.996&  2 &1& 0.43\\
$030201-0027.2$ & 10.311 & 0.049 & 1133 &   3.1062 &  0.118 &  0.052 &  5.505&  2 &1& 0.47\\
$034803-0023.5$ & 10.793 & 0.032 & 1314 &   7.0156 &  0.045 &  0.187 &  6.816&  2 &2& 0.47\\
$044830+0017.9$ & 12.205 & 0.124 & 1457 &   0.2250 &  0.294 &  0.307 &  6.374&  3 &4& 0.49\\
$044944+0056.0$ & 11.447 & 0.086 &  862 &   0.7116 &  0.204 &  0.028 &  6.557&  2 &1& 0.50\\
$045017+0100.7$ & 11.379 & 0.091 &  503 &   0.2056 &  0.224 &  0.288 &  6.294&  2 &4& 0.50\\
$045024+0013.2$ &  9.913 & 0.015 & 1530 &   6.2548 &  0.015 &  0.040 &  3.738&  2 &2& 0.51\\
$045128-0032.7$ &  7.596 & 0.013 & 1534 &   0.7818 &  0.022 &  0.037 &  3.143&  2 &2& 0.51\\
$045206-7043.9$ & 10.542 & 0.086 &  312 &   1.1724 &  0.193 &  0.758 &  6.266&  4 &0& 0.52\\
$045423-7054.1$ & 11.773 & 0.175 &  423 &  34.4540 &  0.566 &  0.385 &  5.068&  5 &6& 0.53\\
$045506-6728.5$ & 12.630 & 0.194 &  267 &  29.8318 &  0.511 &  0.366 &  4.665&  3 &6& 0.53\\
$045511-0101.7$ & 10.247 & 0.023 &  792 &   3.1366 &  0.031 &  0.128 &  7.689&  2 &2& 0.54\\
$045702-6759.7$ & 12.065 & 0.123 &  517 &  45.1273 &  0.322 &  0.279 &  5.124&  2 &6& 0.55\\
$045712-6723.2$ & 12.162 & 0.147 &  273 &  22.7070 &  0.387 &  0.293 &  4.809&  4 &5& 0.55\\
$045720-8023.0$ & 11.615 & 0.228 & 3448 &   0.1835 &  0.721 &  0.368 &  6.321&  6 &4& 0.56\\
$045728-7033.1$ & 12.053 & 0.113 &  676 &   0.8249 &  0.383 &  0.866 &  6.301&  5 &0& 0.56\\
$045750-6957.4$ & 12.354 & 0.147 &  309 &  23.3178 &  0.583 &  0.331 &  5.039&  5 &6& 0.57\\
$045810-6957.0$ & 11.839 & 0.197 &  637 &  39.3889 &  0.575 &  0.385 &  5.162&  5 &6& 0.57\\
$045817-0013.9$ & 10.942 & 0.047 & 1296 &   0.2522 &  0.095 &  0.083 &  6.223&  2 &4& 0.58\\
$045832-7020.8$ & 11.970 & 0.177 &  621 &  35.6997 &  0.607 &  0.435 &  5.332&  5 &6& 0.59\\
$045836-7006.6$ & 12.591 & 0.183 &  295 &  17.2697 &  0.567 &  0.313 &  5.349&  2 &5& 0.59\\
$045914-6935.7$ & 11.061 & 0.100 &  700 &   0.3289 &  0.266 &  0.109 &  5.087&  3 &1& 0.59\\
$045941-6927.4$ & 12.541 & 0.238 &  596 &  31.8223 &  0.833 &  0.463 &  5.099&  4 &6& 0.60\\
$050047-7029.8$ & 10.702 & 0.119 &  706 &   0.1937 &  0.340 &  0.169 &  6.318&  3 &4& 0.60\\
$050327-6909.0$ & 12.473 & 0.172 &  264 &  21.2662 &  0.498 &  0.324 &  5.121&  4 &6& 0.61\\
$050527-6743.2$ &  9.906 & 0.083 &  694 &   2.0243 &  0.240 &  0.364 &  6.509&  4 &3& 0.62\\
$050556-6810.7$ & 11.114 & 0.070 &  391 &   7.3966 &  0.138 &  0.109 &  6.643&  2 &1& 0.62\\
$050558-6810.5$ & 11.220 & 0.075 &  274 &   7.3957 &  0.128 &  0.174 &  7.570&  2 &1& 0.64\\
$050648-7002.2$ & 11.913 & 0.204 &  644 &  47.4545 &  0.643 &  0.414 &  5.557&  4 &6& 0.66\\
$050720-7027.2$ & 12.642 & 0.293 &  528 &  26.3391 &  0.988 &  0.468 &  5.035&  6 &6& 0.66\\
$050818-6846.8$ & 11.497 & 0.106 &  578 &  30.5056 &  0.251 &  0.312 &  5.087&  2 &6& 0.67\\
$050920-7027.4$ & 11.914 & 0.205 &  675 &  37.5444 &  0.636 &  0.436 &  5.369&  5 &6& 0.67\\
$051608-6815.5$ &  9.708 & 0.023 &  699 &   2.1377 &  0.028 &  0.169 &  6.057&  2 &2& 0.67\\
$051833-6813.6$ & 10.089 & 0.176 &  698 &   0.1427 &  0.550 &  0.294 &  6.326&  4 &4& 0.67\\
$052507-6738.6$ & 11.326 & 0.186 & 1152 &  48.0851 &  0.536 &  0.411 &  5.386&  5 &6& 0.67\\
$052557-7011.1$ & 10.556 & 0.059 & 1388 &   1.5803 &  0.132 &  0.010 &  6.235&  2 &1& 0.68\\
$052650-8135.2$ &  7.953 & 0.142 & 2491 &   0.2308 &  0.406 &  0.221 &  6.229&  3 &4& 0.68\\
$052655-6958.9$ & 12.102 & 0.120 &  508 &  28.1222 &  0.316 &  0.421 &  5.134&  2 &6& 0.69\\
$052832-6836.2$ & 10.525 & 0.042 &  700 &   0.2884 &  0.078 &  0.317 &  6.362&  2 &4& 0.71\\
$053014-6926.2$ & 12.456 & 0.153 &  417 &  23.0127 &  0.405 &  0.266 &  4.989&  4 &6& 0.71\\
$053120-7057.5$ & 11.554 & 0.148 &  497 &  52.5205 &  0.438 &  0.356 &  5.384&  3 &6& 0.71\\
$053502-6843.7$ & 11.875 & 0.095 &  669 &   0.5432 &  0.216 &  0.636 &  6.314&  3 &0& 0.72\\
$053936-7958.6$ & 10.418 & 0.040 & 3946 &   0.9208 &  0.114 &  0.410 &  6.262&  6 &0& 0.72\\
$053959-6828.7$ & 10.765 & 0.165 &  693 &   0.1811 &  0.481 &  0.272 &  6.309&  3 &4& 0.73\\
$055122-6812.8$ & 11.747 & 0.120 &  682 &   0.3218 &  0.307 &  0.124 &  5.431&  2 &1& 0.73\\
$055602-0003.8$ & 10.814 & 0.036 & 1374 &   9.8757 &  0.059 &  0.405 &  6.266&  2 &3& 0.73\\
$055624+0013.0$ &  9.471 & 0.014 & 1376 &   1.3206 &  0.013 &  0.567 &  5.726&  2 &2& 0.74\\
$055701+0025.7$ &  9.718 & 0.036 & 1375 &   0.1411 &  0.098 &  0.134 &  6.274&  3 &4& 0.75\\
$055850-0026.5$ & 10.004 & 0.022 & 1375 &   0.5301 &  0.026 &  0.198 &  4.603&  2 &2& 0.75\\ \hline
 \end{tabular}
 \end{center}
\end{table*}
%--------------------------------------------------------------------------

%\newpage
%----------------------------- FIG. 1 -------------------------------------
%--------------------------------------------------------------------------

%----------------------------- FIG. 2 -------------------------------------
%--------------------------------------------------------------------------

%----------------------------- FIG. 3 -------------------------------------
%--------------------------------------------------------------------------

%----------------------------- FIG. 4 -------------------------------------
%--------------------------------------------------------------------------

%%----------------------------- FIG. - ------------------------------------
%\begin{figure*}
%\resizebox{\hsize}{!}{\includegraphics[angle=-90]{dperper2.ps}}
%\caption{\label{fig:dperper2}
%The error on the period as function of the period. The relations is,
%as expected, $\delta P= P^2$.
%}
%\end{figure*}
%%--------------------------------------------------------------------------

%----------------------------- FIG. 5 -----------------------------------
%--------------------------------------------------------------------------

%----------------------------- FIG. 6 ----------------------------------
\begin{figure*}
\includegraphics[width=120mm,angle=-90]{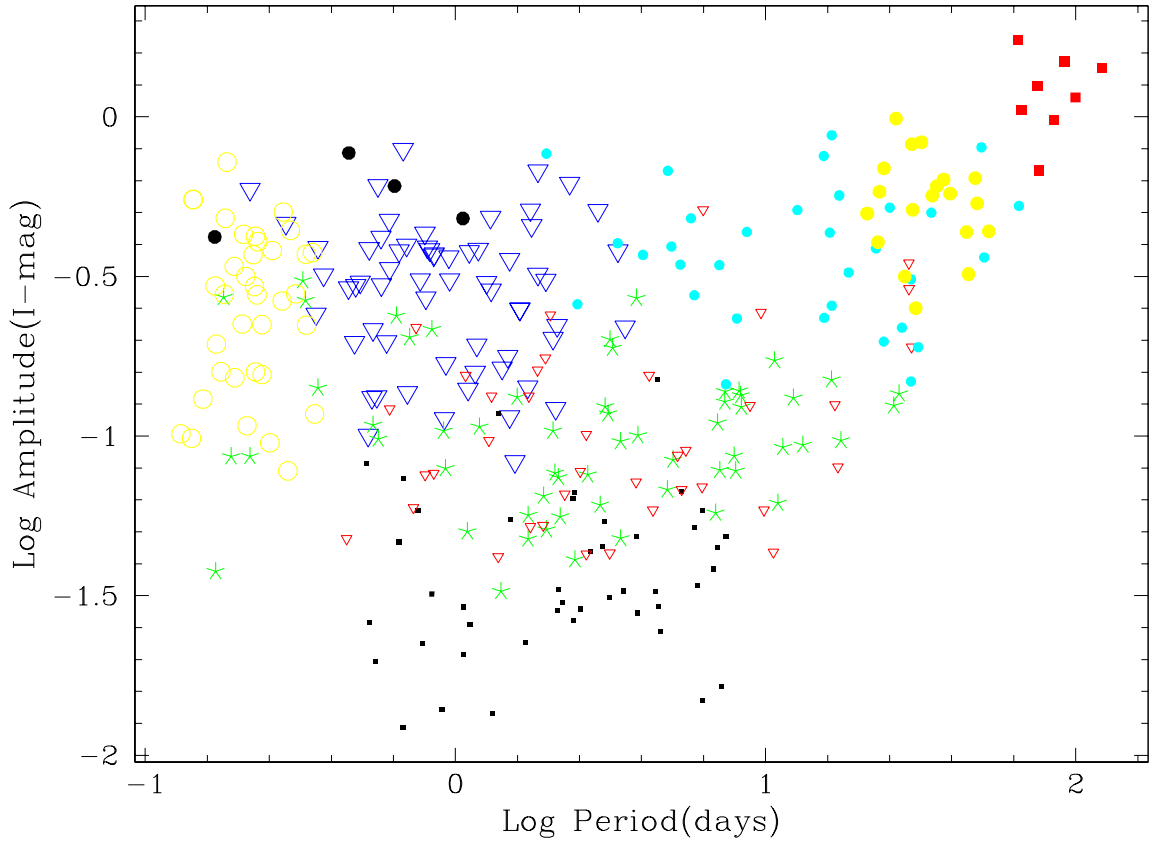}
\caption{\label{fig:classpa} The classification obtained in the
diagram $\log(\mbox{period})$, $\log(\mbox{amplitude})$. Red Giant
stars are excluded from the classification. The Fourier decomposition
is used. The colours description is for the online version.  The
symbols are: Eclipsing binaries: large blue open triangle, open yellow
circles, red small open triangle; RR Lyrae: black large full circles;
Cepheids: yellow small full circles, blue small circles; LPV: red full
large squares; Small amplitude variables: small black
squares. Uncertain cases: green 5 branch star.  }
\end{figure*}
%--------------------------------------------------------------------------

%----------------------------- FIG. 7 -------------------------------------
%--------------------------------------------------------------------------

%----------------------------- FIG. 8 -------------------------------------
\begin{figure*}
\includegraphics[width=120mm,angle=-90]{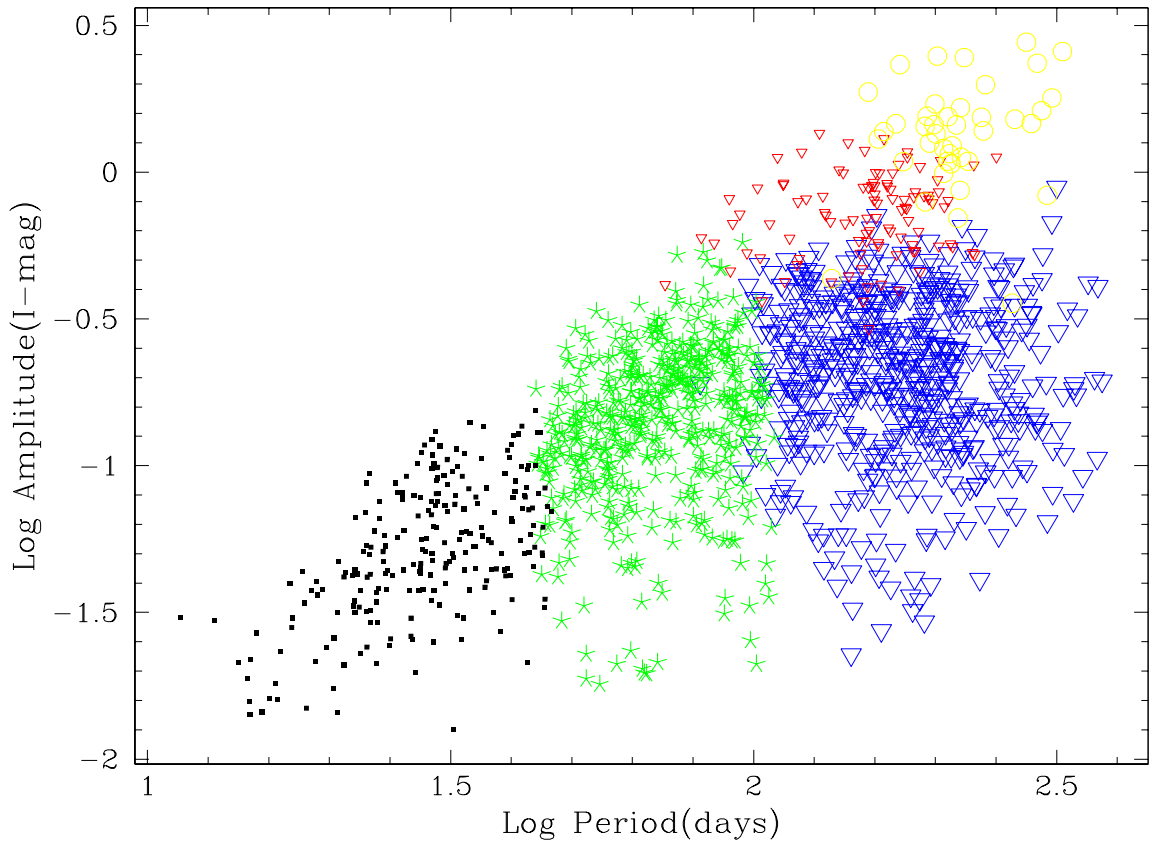}
\caption{\label{fig:classpa}
  The classification obtained in the diagram $\log(\mbox{period})$,
  $\log(\mbox{amplitude})$ for the red giants. The moments of the
  distribution decomposition are used. The classifier forms 5 groups which
  can be identified to three general known variability types: small amplitude
  red variables (small black squares), Semi-regular variable stars
  (5 branch star, open triangles) and Mira stars (open circles).
}
\end{figure*}
%--------------------------------------------------------------------------

%----------------------------- FIG. 9 -------------------------------------
\begin{figure*}
\hspace{1cm}\includegraphics[width=160mm,angle=0]{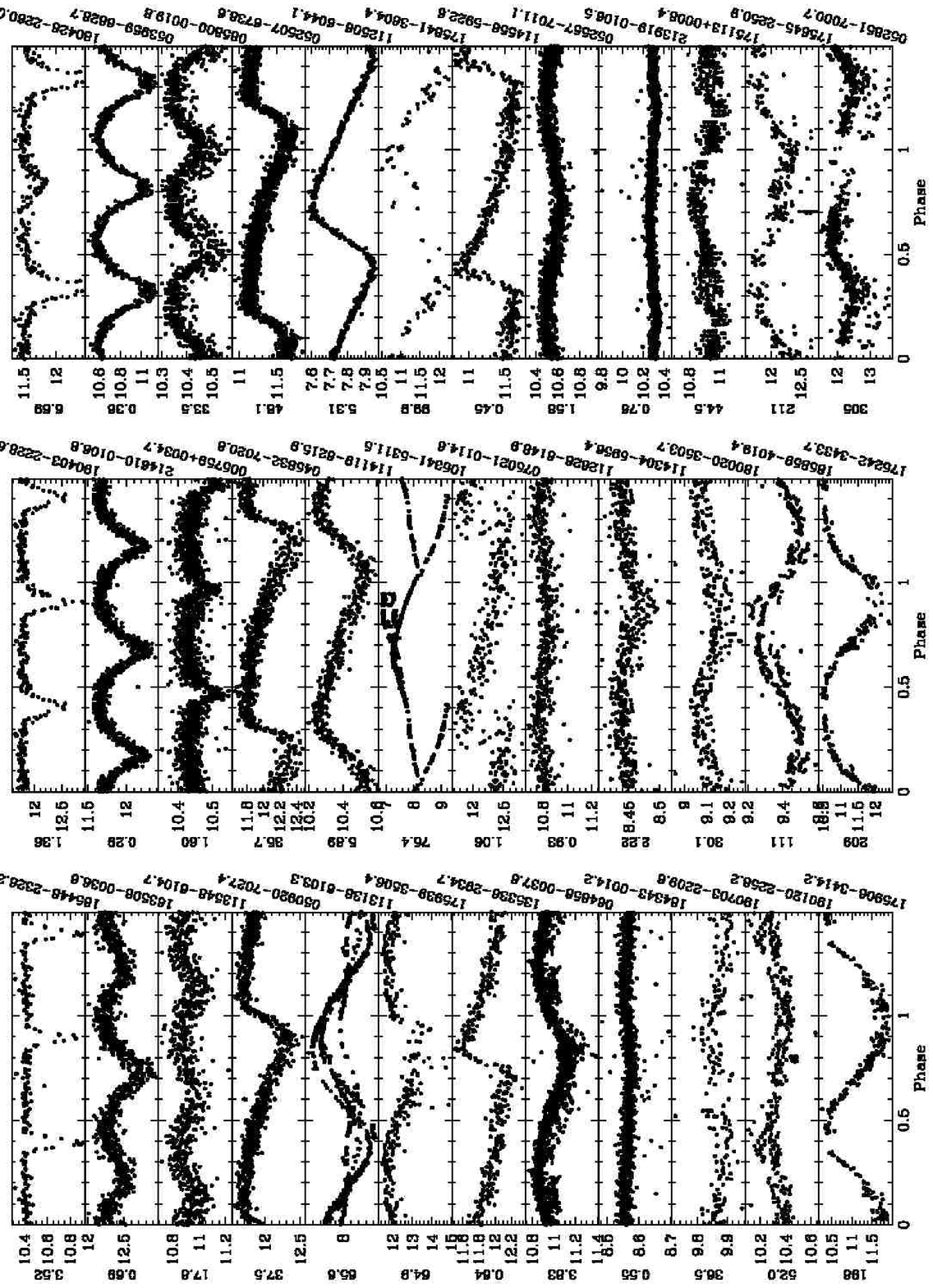}
\caption{\label{fig:samplelc}
  Three examples (on one line) of each of the 12 major classes are
  represented.  We have: 1) eclipsing binaries ($\sim$ EA, EB types),
  2) eclipsing binaries ($\sim$EW types), 3) eclipsing binaries (more
  marginal cases), 4) cepheids, 5) cepheids (with more marginal
  cases), 6) LPVs, 7) RR Lyrae candidates, 8) various case class, 9)
  small amplitude variables, 10) SARVs, 11) SRs, 12) Miras. On the
  right of the folded curve is the ASAS coordinate (equinox 2000), on
  the left is the period in days.
}
\end{figure*}
%--------------------------------------------------------------------------

%#%----------------------------- FIG. 10 ------------------------------------
%#\begin{figure*}
%#\resizebox{\hsize}{!}{\includegraphics[angle=-90]{aitoffnew.ps}}
%#\caption{\label{fig:aitoffnew}
%#Aitoff Projection of the different classes of stars.
%#}
%#\end{figure*}
%#%--------------------------------------------------------------------------
\bsp
\label{lastpage}

\end{document}